\documentclass[aps,prd,preprint,groupedaddress]{revtex4}

\newif\ifpdf
\ifx\pdfoutput\undefined
	\pdffalse
\else
	\pdfoutput=1
	\pdftrue
\fi

\usepackage{amsfonts}
\usepackage{amssymb}
\usepackage{amsmath}

\usepackage{epsfig}
\usepackage{graphicx}

\begin{document}


\title{Galactic Positrons as a Probe of Large Extra Dimensions}


\author{C. Bird}
\affiliation{Department of Physics and Astronomy, University of
Victoria, \\
 Victoria, B.C., V8P 1A1, Canada.}


\date{November 30,2008}

\begin{abstract}
In the last decade, there has been increased interest in the possibility that the Universe contains large additional dimensions. In this article it is proposed that the Kaluza-Klein modes which are predicted to exist in such models will be trapped in the galactic halo, and that the subsequent decays of these modes in the present will generate an observable flux of positrons and $\gamma$-rays from the galactic core. In particular, by restricting the rate of production 511 keV photons to be below that observed by the INTEGRAL satellite, a new set of constraints are imposed on the ADD model.
\end{abstract}

\pacs{}

\maketitle

\par
For the last several years, the idea that spacetime may contain additional dimensions has regained popularity, primarily due to the possibility that such models could resolve the {\it hierarchy problem}\cite{Arkani-Hamed:1998rs,ArkaniHamed:1998nn,Randall:1999vf,Randall:1999ee}.

\par
In these models, the Standard Model fields are trapped on a four-dimensional brane while gravity can propagate in all dimensions. As a result, traditional terrestrial experiments which probe using gauge interactions are unable to constrain these models resulting in the possibility that the extra dimensions could be large. Measurements of the gravitational force at small distances gives an upper limit on the scale of the extra dimensions of $R \lesssim O(0.1 \;mm)$\cite{Adelberger:2003zx,Hoyle:2004cw}.

\par
Although it is difficult to probe these higher dimensions directly, it is possible to search for the effects of Kaluza-Klein modes in astrophysical process such as supernovae and neutron stars \cite{Hannestad:2001jv,Hannestad:2001xi,Hannestad:2003yd}, in the extra-galactic $\gamma$-ray background \cite{Hannestad:2001nq}, or through their effects on Big Bang Nucleosynthesis\cite{Allahverdi:2003aq}. For the purpose of comparison, these results are reviewed in Table \ref{Table::ExtraDimensionBounds}, Table \ref{Table::EDBBN1} and Table \ref{Table::EDBBN2}.

\par
Another possibility is that observable effects could be created in the modern Universe by quasi-stable Kaluza-Klein modes which were produced in the early Universe and which the ADD model predicts will decay in the present at a significant rate. 
\par
In this article I consider the possibility that a significant density of KK gravitons could be trapped in the gravitational potential of the galaxy. Although these modes are quasi-stable, some will decay into electron-positron pairs and into $\gamma$-rays. This is expected to result in an observable flux of $\gamma$-rays both from direct decay of the KK-modes and from subsequent electron-positron annihilation. The observed flux can then be used to constrain the nature and size of the extra dimensions by comparing the predicted flux with the observed flux in the solar system.  
\par
In this article, only the $511 \; keV$ $\gamma$-ray flux produced by positron production in the galaxy will be considered, while the constraints from the full $\gamma$-ray spectrum will be considered in future work. Measurements by the SPI spectrometer on the INTEGRAL satellite \cite{Jean:2003ci,Knodlseder:2005yq} have confirmed previous observations of a flux of $511 \; {\rm keV}$ photons from the galactic center \cite{oldflux1,oldflux2,oldflux3}. These experiments have also determined that the $\gamma$-rays are most likely produced by a diffuse source rather than by a few point sources, which is consistent with a galactic halo composed of dark matter or KK-modes. 
\par
It should also be noted that the constraint considered in this section results from the production of low-energy positrons which lose energy as they travel through the interstellar medium. It is assumed that the positrons become non-relativistic within a short distance and annihilate in the galactic bulge. However the rate of energy loss by these positrons and the cut-off energy above which the positrons can no longer contribute to the observed flux are not well understood, and could potentially cause some uncertainty in the final limits. 

\begin{table}
\begin{center}
\begin{tabular}{|c|c|c|c|c|c|}
\hline
& $d=2$ & $d=3$ &$d=4$&$d=5$& $d=6$ \\ \hline
 $M_*$ & 167 TeV & 21.7 TeV & 4.75 TeV & 1.55 TeV 
& $< 1$  TeV$$
\\ \hline
R& 22 nm & $2.5 \times 10^{-2}$ nm & 1.1 pm & 0.17 pm & $>$ 0.029 pm 
\\ \hline \hline
 $M_*$ & 700  TeV & 25.5  TeV & 2.8 TeV 
& 0.57 TeV & 0.17 TeV \\
\hline R &$1.3 $ nm &$1.9 \times 10^{-2}$ nm &2.4  pm & 0.70 pm &$0.16$ pm
\\ \hline
\end{tabular}
\end{center}
\caption{\label{Table::ExtraDimensionBounds} Bounds on the size of nonwarped extra dimensions and the reduced Planck mass from astrophysical experiments, with the first two lines representing bounds from the $\gamma$-ray background\cite{Hannestad:2001nq} and the second two lines representing bounds from neutron stars\cite{Hannestad:2001xi}. The size of the extra dimensions is given as an upper bound, while the Planck mass given is a lower bound. 
}
\end{table}

\begin{table}
\begin{center}
\renewcommand{\arraystretch}{1.5}
\begin{tabular}{|c|c||c|c|c|c|c|}
\hline 
$m_\phi$ & & $d=2$ & $d=3$ &$d=4$&$d=5$& $d=6$
\\ \hline \hline
1 TeV & $M_*$ & 35 TeV  & 13 TeV & 7.1 TeV & 4.5 TeV & 2.8 TeV
\\ \hline
 & R & $0.52 \mu m$  & 60 pm & 0.59 pm & 39 fm & 7.4 fm
\\ \hline \hline
2 TeV & $M_* $& 47 TeV  & 17 TeV & 9.1  TeV & 5.7 TeV & 3.4 TeV
\\ \hline
 &R & $0.29 \mu m$  & 38 pm & 0.41 pm & 28 fm & 5.7 fm
\\ \hline \hline
$M_*$ &$M_*$& 220 TeV  & 42 TeV &  15 TeV &
7.9 TeV &  4.0 TeV
\\ \hline
&R & $0.013 \mu m$ & 8.5 pm & 0.19 pm &
18 fm &  4.6 fm
\\ \hline
\end{tabular}
\caption{\label{Table::EDBBN1} Lower bounds on the reduced Planck mass and upper bounds on the size of the extra dimensions from nucleosynthesis constraints as a function of the inflaton mass $m_{\phi}$.\cite{Allahverdi:2003aq}}
\end{center}
\end{table}

\section{Kaluza-Klein Modes in the Galaxy}

\par
For the purpose of this calculation, it can be assumed that the $\gamma$-ray flux which results from KK decays depends only on the partial decay width and galactic abundance of the modes. It will also be assumed that all positrons produced in the decay of KK-modes with mass below a certain cut-off, denoted $m_{max}$, will become non-relativistic and annihilate within the galactic bulge. 
The decay widths for graviton decay to electron-positron pairs 
is given in Ref. \cite{Han:1998sg}, 

\begin{equation}\label{Eq:KKwidth}
\Gamma_{e^+e^-} (m_{KK}) = \frac{m_{KK}^3}{80 M_{PL}^2} 
\end{equation}

\noindent
while the abundance of gravitons in the Universe was previously calculated in Ref. \cite{Hall:1999mk} and Ref. \cite{Macesanu:2004gf}. At cosmic scales, the number density of each KK-mode as a function of mass is

\begin{equation}
\begin{split}
n_0(m_{KK}) \simeq & \frac{19 T_0^3}{64 \pi^3 \sqrt{g_*}}  \frac{m_{KK}}{M_{PL}} e^{-t_0/\tau_{KK}} \\ & \times \left( \int_{m_{KK}/T_{RH}}^{\infty} q^3 K_1(q) dq +2 \left(\frac{m_{KK}}{T_{RH}} \right)^{-7} \int_{m_{KK}/T_{MAX}}^{m_{KK}/T_{RH}} q^{10} K_1(q) dq  \right)
\end{split}
\end{equation}


\noindent
where $m_{KK}$ is the mass of the Kaluza-Klein mode,  $T_0$ is the present (neutrino) temperature of the Universe, and $T_{RH}$ is the temperature at which the Universe becomes dominated by radiation.  In this equation, the first term represents KK production by thermal processes which occur during this radiation dominated epoch , while the second term represents KK production during an earlier period of reheating. The second integral also depends on $T_{MAX}$, which is the maximum temperature at which KK modes are produced during the reheating phase. In this section, only the two special cases of $T_{MAX} \sim T_{RH}$ and $T_{MAX} \gg T_{RH}$ will be considered.

\par
The total energy density of Kaluza-Klein modes\index{Kaluza-Klein gravitons} is obtained by summing over all masses. However as before, at the relevant energy scales the difference between the masses of neighbouring KK modes is small and as such the sum can be replaced with an integral over a d-dimensional sphere. The resulting energy density is

\begin{equation}\label{Eq:KKabundance}
\begin{split}
\rho_G = \sum_{all\;modes} m_{KK}n_0(m_{KK}) \simeq & (1.9\times 10^{22} \; GeV^4) S_{d-1}  \left( \frac{T_{RH}}{M_*} \right)^{d+2} I^{(1)}_d (T_{RH}/T_{max})
\end{split}
\end{equation}

\noindent
where

\begin{equation} \label{Eq:SpInt}
\begin{split}
I_d^{(\sigma)} (\beta ) = \int_{0}^{\infty} dz e^{-t_0/\tau_{KK}} & \left(  \; z^{d+\sigma}   \int_{z}^{\infty} dq \; q^3 K_1(q) \right. \left. + 2 \; z^{d+\sigma-7} \int_{\beta z}^{z} dq \; q^{10} K_1(q) \right)
\end{split}
\end{equation}

\noindent
represents the integral over all modes. 

\par
The bound derived in this section assumes that these KK modes, which were formed in the early Universe, have become trapped in the gravitational potential of the galaxy. Using the cosmological KK-mode abundance, 
the distribution of KK modes in the galaxy can then be approximated as

\begin{equation}\label{Eq:KKdist}
\rho_{KK}(r) = \frac{\rho_G}{\rho_{DM,cosmic}} \rho_{DM}(r)
\end{equation}

\noindent
where $\rho_{DM,cosmic}$ is the total cosmic dark matter abundance, and

\begin{equation}
\rho_{DM} (r) = \rho_0 exp \left( -\frac{2}{\alpha} \left[ \left(\frac{r}{r_0} \right)^{\alpha} - 1 \right] \right)
\end{equation}

\noindent
with $r_0 = 20h^{-1} kpc$ and $0.1 < \alpha < 0.2$,is the galactic dark matter distribution \cite{Navarro:2003ew}. The constant $\rho_0$ is determined by requiring $\rho_{DM} (8.5 kpc) = 0.3 \; GeV/cm^3$, which is the accepted value of the dark matter density at the Solar system. In this calculation it is assumed that the KK modes are distributed in the halo with the same mass distribution as in the early Universe. In practice the lightest modes are too light to be captured in the halo, however the effect of these missing modes is small. 

\section{The 511 keV $\gamma$-ray Flux}
\par
\begin{figure}
\begin{center}
\psfig{file=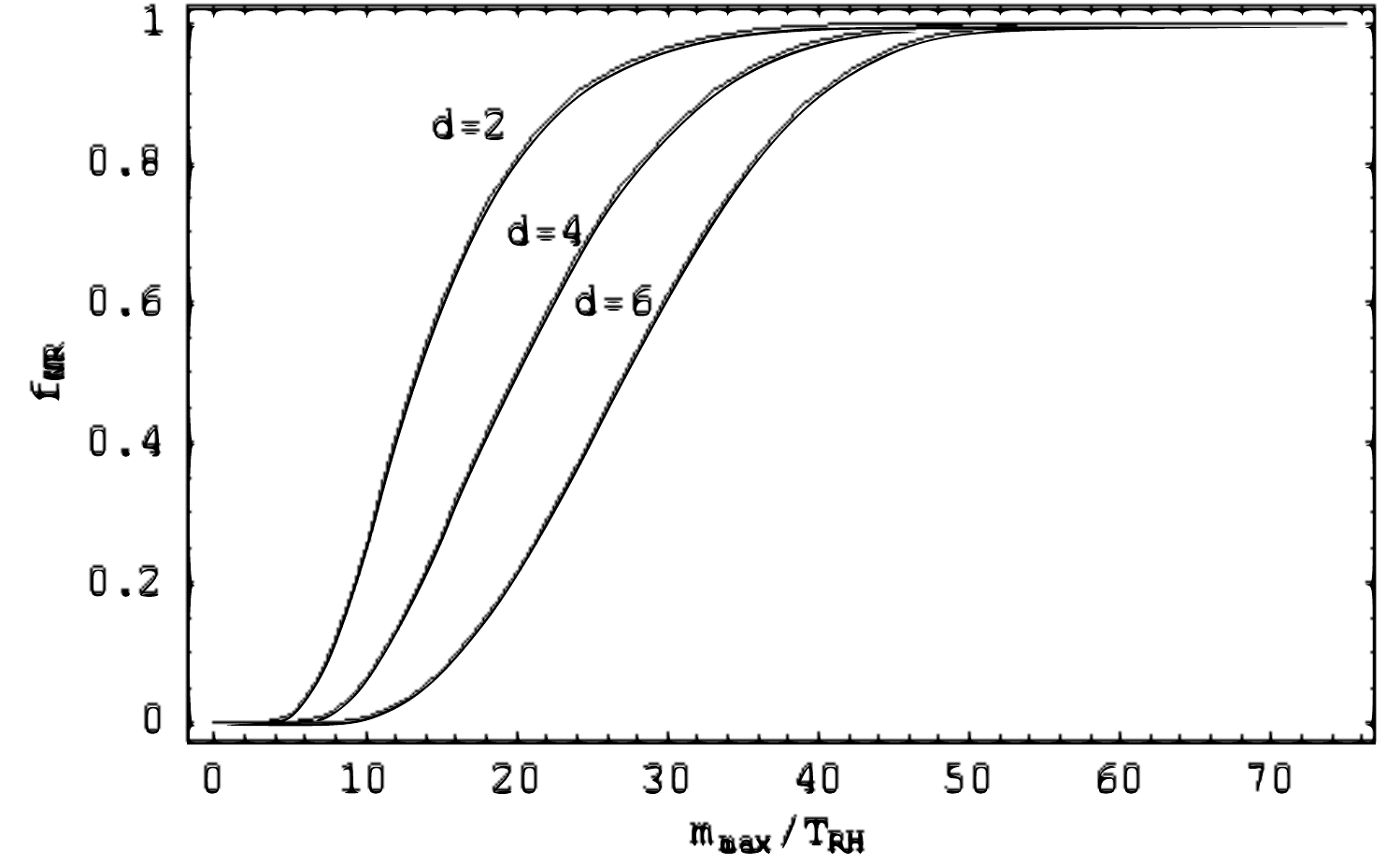,width=\textwidth,angle=0} \\ (a) \\
\psfig{file=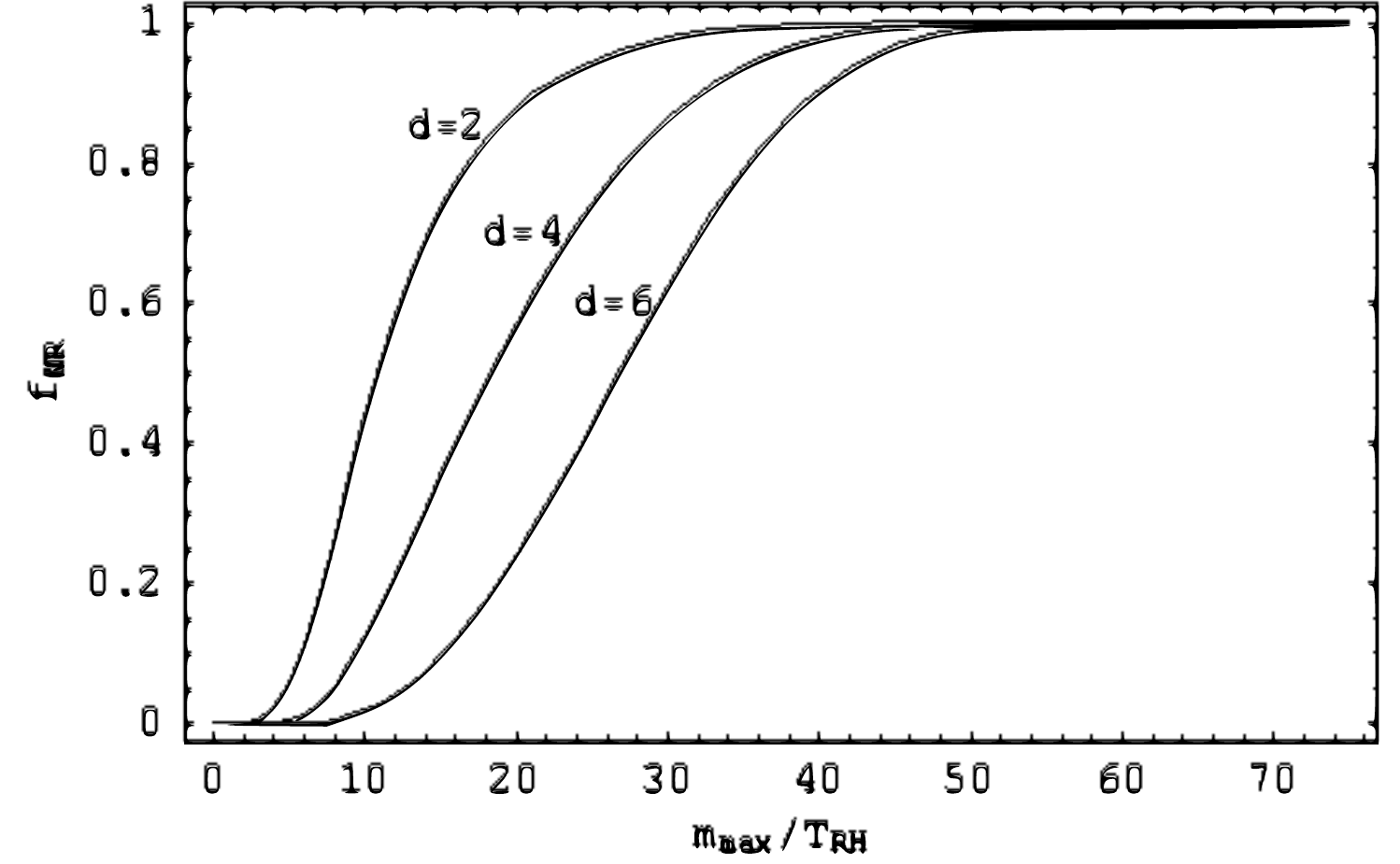,width=\textwidth,angle=0} \\ (b)
\end{center}
\caption{\label{Fig:PosFrac}The fraction of Kaluza-Klein modes with mass below $m_{max}$ for the cases of (a) $T_{MAX} \sim T_{RH}$ and (b) $T_{MAX} \gg T_{RH}$.}
\end{figure}

Using similar methods, the rate of positron production in the Universe can be determined by summing over the partial width for the decay $KK \to e^+e^-$ of each KK mode, 

\begin{equation}
\begin{split}
N_{e^+,cosmic}& =\sum_{all \; modes} \Gamma_{e^+e^-}(m_{KK})n_0(m_{KK}) \\& \simeq  (1.9\times 10^{22} \; GeV^4) S_{d-1}  \left( \frac{T_{RH}}{M_*} \right)^{d+2} \frac{T_{RH}^2}{80 M_{PL}^2} I^{(3)}_d (T_{RH}/T_{max})
\end{split}
\end{equation}



\noindent
and the number density of non-relativistic positrons produced per unit time in the galaxy follows from Eq \ref{Eq:KKdist},

\begin{equation} \label{Eq:PosDen}
\begin{split}
N_{511 keV} (r)  & \simeq N_{e^+,cosmic} \frac{ \rho_{DM}(r)}{\rho_{DM,cosmic}}\\&\simeq  (2.57 \times 10^{-3d - 4} \; cm^{-3} s^{-1} )(0.85 + 1.49 \alpha) S_{d-1}  \left( \frac{1000 \; T_{RH}}{M_*} \right)^{d+2}  \\ & \times \left(\frac{T_{RH}}{1 \; MeV}\right)^2 I_d^{(3)}(T_{RH}/T_{MAX}) f_{NR}(m_{max}/T_{RH}) exp \left( -\frac{2}{\alpha} \left[ \left(\frac{r}{r_0} \right)^{\alpha} - 1 \right] \right)
\end{split}
\end{equation}


\noindent
where the integral $I_d^{(3)}$ is given in Eq \ref{Eq:SpInt}, and $f_{NR}(m_{max}/T_{RH})$ is the fraction of Kaluza-Klein modes which are lighter than $m_{max}$, and which are assumed to decay to non-relativistic positrons. This fraction is plotted in Figure \ref{Fig:PosFrac} for the cases of $T_{MAX} \sim T_{RH}$ and $T_{MAX} \gg T_{RH}$.

\par
As outlined in Ref \cite{Picciotto:2004rp}, the resulting flux of $511 \; keV$ $\gamma$-rays at the Solar system can be derived by calculating the line-of-sight integral of this density. However it should be noted that not all positrons contribute to the $511 \; keV$ $\gamma$-ray flux. As outlined in Ref \cite{positronium1,positronium2,Beacom:2005qv} positrons produced in the galaxy can form positronium states with electrons before annihilating. The positronium states only produce $511 \; keV$ photons in approximately $25 \%$ of the annihilations, while the other states decay to three photons each with lower energies. By measuring the flux of $511 \; keV$ photons relative to lower energy photons, the fraction of positrons bound into positronium states in the galactic core is determined to be $f = 0.967 \pm 0.022$ \cite{Jean:2005af}. As a result, on average each positron produced contributes $2(1-0.75 f) = 0.55$ photons to the $511 \; keV$ flux. Therefore the rate of positron production is expected to be a factor of $\sim 3.6$ larger than would be expected based solely on the $511 \; keV$ flux.

\par
As discussed previously, the $511 \; keV$ flux has been measured by several experiments. The SPI spectrometer on the INTEGRAL $\gamma$-ray observatory measured
 an azimuthally symmetric distribution of $511\; keV$ photons with FWHM
 of $\sim 8^\circ \pm 1^\circ$, and total flux $\Phi_{obs} = (1.05\pm 0.06) \times 10^{-3} \; ph \; cm^{-2} \; s^{-1}$ \cite{Jean:2003ci,Knodlseder:2005yq}. Integrating the density in Eq \ref{Eq:PosDen} over
 this solid angle with the line of sight integral gives a total predicted flux of

\begin{equation}
\begin{split}
\Phi_{th}&= (0.86 \times 10^{20-3d} \; cm^{-2} s^{-1} )
(1.4 - 3.8 \alpha) S_{d-1}  \left( \frac{1000 \; T_{RH}}{M_*} \right)^{d+2}  \\ & \times \left(\frac{T_{RH}}{1 \; MeV}\right)^2 f_{NR}(m_{max}/T_{RH}) I_d^{(3)}(T_{RH}/T_{MAX}) 
\end{split}
\end{equation}


\noindent
Comparing this result with the observed $511 \;keV$ $\gamma$-ray flux constrains the Planck mass,

\begin{equation}\label{DMC}
\begin{split}
\left( \frac{M_*}{1000 \; T_{RH}} \right)^{d+2} \gtrsim (0.43 \pm 0.03) \times 10^{23-3d} &  
(1.4 - 3.8 \alpha) S_{d-1} \left(\frac{T_{RH}}{1 \; MeV}\right)^2 \\ & \times f_{NR}(m_{max}/T_{RH})I_d^{(3)}(T_{RH}/T_{MAX}) 
\end{split}
\end{equation}


\noindent
where as before $S_d = 2 \pi^{d/2} / \Gamma(d/2)$, and where it is assumed that the contribution from direct production of $511 \; keV$ photons is negligible. Although it is possible to produce a flux of $\gamma$-rays in this energy range through direct production, the effect is expected to be small and will not significantly alter the constraints given.



\begin{table}
\begin{center}
\begin{tabular}{|c|c|c|c|c|c|}
\hline &d=2&d=3&d=4&d=5&d=6\\ \hline 
$T_{MAX} \gg T_{RH}$ &&&&& \\ \hline
$\alpha=0.1$ & 580 \; TeV & 20 \; TeV & 2.1 \; TeV & 0.43 \; TeV & 0.13 \; TeV \\ 
$\alpha=0.2$ & 520 \; TeV & 18 \; TeV & 1.9 \; TeV & 0.39 \; TeV & 0.12 \; TeV \\ \hline
$T_{MAX} \sim T_{RH}$ &&&&& \\ \hline
$\alpha=0.1$ & 340 \; TeV & 12 \; TeV & 1.2 \; TeV & 0.17 \; TeV & 0.069 \; TeV \\
 $\alpha=0.2$ & 300 \; TeV & 11 \; TeV & 1.1 \; TeV & 0.16 \; TeV & 0.066 \; TeV \\ \hline
\end{tabular}
\caption{\label{Table::PosBound}Limits of sensitivity on $M_*$ (TeV) for different values of $\alpha$ and different dimensions, assuming $T_{RH} = 1 \; MeV$. For larger values of $M_*$, the positron flux from decaying KK modes is lower than the observed $511 \; keV$ $\gamma$-ray flux. }
\end{center}
\end{table}

\begin{table}
\begin{center}
\begin{tabular}{|c|c|c|c|c|c|}
\hline &d=2&d=3&d=4&d=5&d=6\\ \hline 
$T_{MAX} \gg T_{RH}$ &&&&& \\ \hline
$\alpha=0.1$ & 1.9 \; nm & 30 \; pm & 3.7 \; pm & 1.0 \; pm & 0.45 \; pm\\ 
$\alpha=0.2$ & 2.3 \; nm & 35 \; pm& 4.3 \; pm& 1.2 \; pm& 0.50 \; pm\\ \hline
$T_{MAX} \sim T_{RH}$ &&&&& \\ \hline
$\alpha=0.1$ & 5.5 \; nm & 69 \; pm & 8.6 \; pm & 3.8 \; pm & 1.0 \; pm\\
 $\alpha=0.2$ & 7.0 \; nm& 79 \; pm & 9.7 \; pm& 4.1 \; pm& 1.1 \; pm\\ \hline
\end{tabular}
\caption{\label{Table:PosBound2} Limits of sensitivity on the size of extra dimensions, R  for different values of $\alpha$ and different dimensions and assuming $T_{RH} = 1 \; MeV$. For smaller values of R, the positron flux from decaying KK modes is lower than the observed $511 \; keV$ $\gamma$-ray flux.}
\end{center}
\end{table}

\par
 The corresponding bounds on the Planck mass and the size of the extra dimensions are given in Table \ref{Table::PosBound} and Table \ref{Table:PosBound2} respectively. In each bound, it is assumed that $T_{RH} \sim 1 \; MeV$, although recent data from the CMB and from large scale structure suggest $T_{RH} > 4 \; MeV$ for most models \cite{Hannestad:2004px}.  
\nocite{Kawasaki:1999na,Kawasaki:2000en} If this stronger temperature bound is used, then each of the constraints on $M_*$ is improved by a factor of $4^{(d+4)/(d+2)}$.  The corresponding bounds
 in this case are given in Table \ref{Table::PosBound3} and Table \ref{Table:PosBound4} respectively.

\begin{table}
\begin{center}
\begin{tabular}{|c|c|c|c|c|c|}
\hline &d=2&d=3&d=4&d=5&d=6\\ \hline 
$T_{MAX} \gg T_{RH}$ &&&&& \\ \hline
$\alpha=0.1$ & 4600 \; TeV & 140 \; TeV & 13 \; TeV & 2.6 \; TeV & 0.74 \; TeV\\ 
$\alpha=0.2$ & 4200 \; TeV & 130 \; TeV & 12 \; TeV & 2.3 \; TeV & 0.68 \; TeV \\ \hline
$T_{MAX} \sim T_{RH}$ &&&&& \\ \hline
$\alpha=0.1$ & 2700 \; TeV & 83 \; TeV & 7.8 \; TeV & 1.0 \; TeV & 0.39 \; TeV \\
 $\alpha=0.2$ & 2400 \; TeV & 77 \; TeV & 7.0 \; TeV & 0.95 \; TeV & 0.37 \; TeV \\ \hline
\end{tabular}
\caption{\label{Table::PosBound3}Limits of sensitivity on $M_*$ (TeV) for different values of $\alpha$ and different dimensions, assuming $T_{RH} \sim 4 \; MeV$ and $f_{NR} \approx 1$.  }
\end{center}
\end{table}

\begin{table}
\begin{center}
\begin{tabular}{|c|c|c|c|c|c|}
\hline &d=2&d=3&d=4&d=5&d=6\\ \hline 
$T_{MAX} \gg T_{RH}$ &&&&& \\ \hline
$\alpha=0.1$ & 30 \; pm & 1.2 \; pm & 0.23 \; pm & 0.082 \; pm & 0.045 \; pm\\ 
$\alpha=0.2$ & 36 \; pm & 1.4 \; pm& 0.27 \; pm& 0.098 \; pm& 0.050 \; pm\\ \hline
$T_{MAX} \sim T_{RH}$ &&&&& \\ \hline
$\alpha=0.1$ & 86 \; pm & 2.7 \; pm & 0.54 \; pm & 0.31 \; pm & 0.10 \; pm\\
 $\alpha=0.2$ & 110 \; pm& 3.1 \; pm & 0.61 \; pm& 0.34 \; pm& 0.11 \; pm\\ \hline
\end{tabular}
\caption{\label{Table:PosBound4} Limits of sensitivity on the size of extra dimensions, R  for different values of $\alpha$ and different dimensions assuming $T_{RH} \sim 4 \; MeV$ and $f_{NR} \approx 1$. }
\end{center}
\end{table}

\par
It is also assumed that the mass at which decaying modes no longer produce non-relativistic positrons is sufficiently high as to not affect this bound. The production of nonrelativistic positrons requires $m_{max} \lesssim O(200 \; MeV)$ \cite{Boehm:2003bt}. Limits on $\gamma$-ray production by bremsstrahlung processes suggest $m_{max} \lesssim 40 \; MeV$ \cite{Beacom:2004pe} or $m_{max} \lesssim 6 \; MeV$ \cite{Beacom:2005qv}. If these bounds are used as a mass cut-off, the resulting constraints on the ADD model are weaker than from other astrophysics experiments. However it should be noted that these bounds assume all positrons are created with uniform energy. In this model, the positrons have a range of energies corresponding to the range of KK-mode masses, and as such $m_{max}$ can be larger without violating these bounds. As indicated in Figure \ref{Fig:PosFrac}, $m_{max}/T_{RH}$ can be as low as $\sim 40$ without significantly weakening the constraints.

%

\par
It should also be noted that these bounds are based on several assumptions. First, there is an assumption that the relative abundance of KK modes\index{Kaluza-Klein gravitons} of different masses is the same in the galaxy as in the early Universe. There are also uncertainties in the dark matter halo profile of the galaxy, which can affect the observed flux, although this variation is small for most reasonable profiles and parameters. It is also assumed that the positrons annihilate within a short distance compared to the scale of the galactic center. This diffusion process is not completely understood, and if the positrons travel further then the observed flux would have a wider angular distribution which could improve these bounds \cite{Beacom:2004pe}. It is also assumed that all annihilating positrons produce $511 \; keV$ photons (or lower energy $\gamma$-rays from the decays of positronium states.). However if a significant fraction of the positrons annihilate at higher energies, these photons would not be counted in the observed $511 \; keV$ flux. This would result in less stringent constraints on the number of positrons, and would weaken the bounds given in Table \ref{Table::PosBound} and Table \ref{Table:PosBound2}, while improving bounds obtained from considering the entire $\gamma$-ray spectrum.


\section{Summary}

\par
 From Table \ref{Table::PosBound}, it is apparent that the accumulated KK modes in the galactic core provide a strong bound on the size of the extra dimensions in the ADD model,as well as providing an explanation for the source of the observed population of galactic positrons. In addition, other models containing extra dimensions could be constrained in a similar manner. 
 \par
 When these modes do decay, they inject photons and positrons into the galaxy which could be observed by existing experiments. As an example, the non-relativistic positron density in the galaxy which is produced by the decay of light KK-modes was derived. By requiring the flux of $511 \; keV$ photons produced by the subsequent annihilations of the positrons to be lower than the observed flux, additional constraints have been placed on the properties of extra dimensions.

\par
In summary, Kaluza-Klein gravitons which were produced in the early Universe can still exist in the present. Furthermore, these particles can accumulate in the galactic halo, leading to a high density in the galactic core. If KK modes heavier than $\sim 40 \; MeV$ can exist in the galaxy with a significant abundance and can produce nonrelativistic positrons, then the bounds from galactic positron production are significantly stronger than the bounds from collider experiments for $d \leq 4$, and are comparable to the bounds from the extragalactic $\gamma$-ray background and other astrophysics experiments for all dimensions.

\bibliography{gpex}

\end{document}